
\catcode`@=11

\newskip\ttglue

\font\twelverm=cmr12 \font\twelvebf=cmbx12
\font\twelveit=cmti12 \font\twelvesl=cmsl12

\font\ninerm=cmr9
\font\eightrm=cmr8
\font\sixrm=cmr6
\font\eighti=cmmi8   \skewchar\eighti='177
\font\sixi=cmmi6     \skewchar\sixi='177
\font\ninesy=cmsy9   \skewchar\ninesy='60
\font\eightsy=cmsy8  \skewchar\eightsy='60
\font\sixsy=cmsy6    \skewchar\sixsy='60
\font\eightbf=cmbx8
\font\sixbf=cmbx6
\font\eighttt=cmtt8  \hyphenchar\eighttt=-1
\font\eightit=cmti8
\font\eightsl=cmsl8

\def\smalltype{\def\rm{\fam0\eightrm}
 			\textfont0=\eightrm  \scriptfont0=\sixrm  \scriptscriptfont0=\fiverm
 			\textfont1=\eighti   \scriptfont1=\sixi   \scriptscriptfont1=\fivei
 			\textfont2=\eightsy  \scriptfont2=\sixsy  \scriptscriptfont2=\fivesy
 			\textfont3=\tenex    \scriptfont3=\tenex  \scriptscriptfont3=\tenex
    \textfont\itfam=\eightit  \def\it{\fam\itfam\eightit}
	   \textfont\slfam=\eightsl  \def\sl{\fam\slfam\eightsl}
	   \textfont\ttfam=\eighttt  \def\tt{\fam\ttfam\eighttt}
    \textfont\bffam=\eightbf  \scriptfont\bffam=\sixbf
        \scriptscriptfont\bffam=\fivebf  \def\bf{\fam\bffam\eightbf}
    \tt  \ttglue=.5em plus.25em minus.15em
    \normalbaselineskip=9pt
    \setbox\strutbox=\hbox{\vrule height7pt depth2pt width0pt}
    \let\sc=\sixrm  \let\big=\eightbig  \normalbaselines\rm}
\def\eightbig#1{{\hbox{$\textfont0=\ninerm\textfont2=\ninesy
    \left#1\vbox to6.5pt{}\right.\n@space$}}}

\def\medtype{\def\rm{\fam0\tenrm}
 			\textfont0=\tenrm  \scriptfont0=\sevenrm  \scriptscriptfont0=\fiverm
 			\textfont1=\teni   \scriptfont1=\seveni   \scriptscriptfont1=\fivei
 			\textfont2=\tensy  \scriptfont2=\sevensy  \scriptscriptfont2=\fivesy
 			\textfont3=\tenex    \scriptfont3=\tenex  \scriptscriptfont3=\tenex
    \textfont\itfam=\tenit  \def\it{\fam\itfam\tenit}
	   \textfont\slfam=\tensl  \def\sl{\fam\slfam\tensl}
	   \textfont\ttfam=\tentt  \def\tt{\fam\ttfam\tentt}
    \textfont\bffam=\tenbf  \scriptfont\bffam=\sevenbf
        \scriptscriptfont\bffam=\fivebf  \def\bf{\fam\bffam\tenbf}
    \tt  \ttglue=.5em plus.25em minus.15em
    \normalbaselineskip=12pt
    \setbox\strutbox=\hbox{\vrule height8.5pt depth3.5pt width0pt}
    \let\sc=\eightrm  \let\big=\tenbig  \normalbaselines\rm}

\def\bigtype{\let\rm=\twelverm \let\bf=\twelvebf
\let\it=\twelveit \let\sl=\twelvesl \rm}

\def\footnote#1{\edef\@sf{\spacefactor\the\spacefactor}#1\@sf
    \insert\footins\bgroup\smalltype
    \interlinepenalty100 \let\par=\endgraf
    \leftskip=0pt  \rightskip=0pt
    \splittopskip=10pt plus 1pt minus 1pt \floatingpenalty=20000
  \vskip4pt\noindent\hskip20pt\llap{#1\enspace}
\bgroup\strut\aftergroup\@foot\let\next}
\skip\footins=12pt plus 2pt minus 4pt \dimen\footins=30pc

\def\bigfont{\magnification=1200 \baselineskip=20pt}

  \def\d{\delta}
 \def\g{\gamma} 
  \def\o{\omega}
  \def\O{\Omega}

\def\cl#1{\centerline{#1}}
\def\clbf#1{\centerline{\bf #1}}

\def\is#1{{\narrower\smallskip\noindent#1\smallskip}}

\long\def\myname{\medskip
\cl{Kiho Yoon}
\cl{Department of Economics, Korea University}
\cl{145 Anam-ro, Seongbuk-gu, Seoul, Korea 02841}
\cl{ \tt kiho@korea.ac.kr}
\cl{\tt http://econ.korea.ac.kr/\~{ }kiho}
\medskip}

\def\ve{\vfill\eject}

\def\frac#1#2{{#1 \over #2}}
\def\Re{I\!\!R}

\newcount\sectnumber
\def\Section#1{\global\advance\sectnumber by 1 \bigskip
           \noindent{\bigtype {\bf \the\sectnumber  \ \ \ #1}} \medskip}

\def\defn#1{\medskip\noindent {\bf Definition #1. }}

\def\lemma#1{\medskip\noindent {\bf Lemma #1.} \it}
\def\thm#1{\medskip\noindent {\bf Theorem #1.} \it}
\def\corr#1{\medskip\noindent {\bf Corollary #1.} \it}

\def\eg#1{\medskip\noindent {\bf Example #1.}}

\def\ok{\smallskip \rm}

\def\pf{\medskip\noindent Proof: \/}
\def\pfo#1{\medskip\noindent {\bf Proof of #1:\/}}
\def\endpf{\hfill {\it Q.E.D.} \smallskip}

\def\appx{\bigskip {\bigtype \clbf{Appendix}} \medskip}

\newcount\notenumber
\def\note#1{\global\advance\notenumber by 1
            \footnote{$^{\the\notenumber}$}{#1}\tenrm}

\def\ref{\bigskip \centerline{\bf REFERENCES} \medskip}

\def\emet{{\it Econometrica\/ }}
\def\jet{{\it Journal of Economic Theory\/ }}

\def\geb{{\it Games and Economic Behavior\/ }}

\def\jpube{{\it Journal of Public Economics\/ }}

\def\et{{\it Economic Theory\/ }}

\def\te{{\it Theoretical Economics\/ }}

\def\wp{{\it working paper\/ }}

\def\paper#1#2#3#4#5{\noindent\hangindent=20pt#1 (#2), ``#3,'' #4, #5.\par}
\def\wp#1#2#3#4{\noindent\hangindent=20pt#1 (#2), ``#3,'' #4.\par}
\def\book#1#2#3#4{\noindent\hangindent=20pt#1 (#2), {\it #3,} #4.\par}


\bigfont
\def\th{\theta}
\def\Th{\Theta}

{ \ \ }

\vskip 1cm

{\bigtype
\clbf{The uniqueness of dynamic Groves mechanisms}
\clbf{on restricted domains\footnote*{I thank Ruggiero Cavallo for clarifying the main contribution of the paper. This work was supported by a Korea University Grant (K1911191).}}
}

\vskip 1cm
\bigskip

\myname

\vskip 0.5cm

\vskip 0.5cm

\clbf{Abstract}
\is{\baselineskip=12pt This paper examines necessary and sufficient conditions for the uniqueness of dynamic Groves mechanisms when the domain of valuations is restricted. Our approach is to appropriately define the total valuation function, which is the expected discounted sum of each period's valuation function from the allocation and thus a dynamic counterpart of the static valuation function, and then to port the results for static Groves mechanisms to the dynamic setting.}


\is{\baselineskip=12pt Keywords: Groves mechanism, dynamic mechanism design, ex-post incentive compatibility, outcome efficiency}

\ve

\Section{Introduction}

The Groves mechanisms hold enormous theoretical value in mechanism design theory. They are outcome efficient and dominant strategy incentive compatible, that is, they maximize the sum of players' valuations from the allocation and induce truth-telling as a dominant strategy equilibrium. Moreover, Green and Laffont (1977), Walker (1978),  Holmstr\"om (1979) and Carbajal (2010) have shown that every outcome efficient and dominant strategy incentive compatible mechanism is a Groves mechanism. In particular, Green and Laffont (1977) have proven the uniqueness result both for unrestricted domains and for a restricted domain containing continuous valuation functions, whereas Walker (1978) has proven the uniqueness result for the class of concave valuation functions on a convex subset of a Euclidean space. Holmstr\"om (1979) has subsequently established a result for any smoothly connected domain of valuation functions, which thus implies the previous results. Carbajal (2010) has provided a necessary and sufficient condition for the uniqueness, which naturally implies most of the previous results.\note{With certain assumptions on the domain, Williams (1999) and Krishna and Perry (2000) have proven the result that every outcome efficient and Bayesian incentive compatible mechanism is payoff-equivalent to some Groves mechanism from an interim perspective.}  Therefore, when searching for mechanisms that additionally satisfy other desirable properties, such as individual rationality and budget balance, we can restrict our attention to Groves mechanisms.\note{Some authors use the term `efficiency' to mean outcome efficiency. Others use efficiency to mean outcome efficiency plus budget balance. We follow the latter convention. Thus, a mechanism is efficient when it maximizes the sum of players' valuations from the allocation and it does not run a monetary deficit.}

Given the importance of the uniqueness result, it is an interesting research agenda to extend the uniqueness of Groves mechanisms to dynamic environments in which players' private information evolves over time and decisions are made in each period.\note{See Bergemann and Said (2010), Vohra (2012), Pavan (2017) and Bergemann and V\"alim\"aki (2019) for surveys on dynamic mechanism design literature.} Cavallo (2008) has extended the uniqueness result for unrestricted domains, following the method of proof in Green and Laffont (1977). The purpose of the present paper is to investigate whether the uniqueness of Groves mechanisms in dynamic environments continues to hold when the domain is further restricted.

In the next section, we carefully describe dynamic environments, and then define the class of dynamic Groves mechanisms, which is a dynamic version of the class of static Groves mechanisms, for Markovian environments and show that they are outcome efficient and periodic ex-post incentive compatible. The class of dynamic Groves mechanisms encompasses the dynamic pivot mechanism of Bergemann and V\"alim\"aki (2010) as well as the team mechanism of Athey and Segal (2013), two of the most famous mechanisms in the dynamic mechanism design literature. In Section 3, we specify conditions for the restricted domains of valuations and examine necessary and sufficient conditions for the uniqueness of dynamic Groves mechanisms. Our approach is to appropriately define the {\it total valuation function\/}, which is the expected discounted sum of each period's valuation function from the allocation and thus a dynamic counterpart of the static valuation function, and then to port the results for static Groves mechanisms to the dynamic setting. In particular, we port the recent results of Carbajal (2010) which provide a necessary and sufficient condition for the uniqueness of Groves mechanisms as well as a sufficient condition on restricted domains. The final section contains a brief summary.

The uniqueness of dynamic Groves mechanisms can be interpreted as a characterization of payoff equivalence for outcome efficient and periodic ex-post efficient dynamic mechanisms. The payoff equivalence result is well-established in mechanism design literature: Representative works include, besides the papers mentioned above, Krishna and Maenner (2001) and Milgrom and Segal (2002) for static environments and Pavan {\it et al.\/} (2014), Skrzypacz and Toikka (2015) and Bergemann and Strack (2015) for dynamic environments. Compared to the payoff equivalence results for restricted domains in the dynamic mechanism design literature, the current paper deals with multi-dimensional type spaces as well as periodic ex-post incentive compatibility.\note{Other papers cited in this paragraph consider perfect Bayesian incentive compatibility and assume that the type space is one-dimensional in each period, mainly due to their interest in revenue maximization problem. Needless to say, these papers differ in other respects and  also they are more general than the current paper in other aspects. Please refer to the original papers for details. Please refer also to Carbajal (2010) for a discussion on the significance of his results in comparison to other payoff equivalence results in static mechanism design.}

\Section{The Dynamic Groves Mechanism}

\noindent {\sl 2.1. The environment}
\medskip

There is a set $I = \{1,\ldots ,n\}$ of players and a countably infinite number of periods, indexed by  $t \in \{0,1,\ldots\}$. Player $i$'s type in period $t$ is $\th_i^t \in \Th_i$. We assume that this is private information. Let $\th^t = (\th_1^t, \ldots, \th_n^t)$ and $\Th = \prod_{i =1}^n \Th_i$.\note{We may include public information, say $\th_0^t \in \Th_0$, to be more realistic. We dispense with this additional notation for clearer presentation of the main idea.} We assume that $\Th$ is a Borel space, i.e., a Borel subset of a complete and separable metric space. Let ${\cal B}(\Th)$ be the Borel $\sigma$-algebra on $\Th$. After $\th^t \in \Th$ is realized in period $t$, a public action $a^t \in A$ is determined. We assume that $A$ is a Borel space, with the Borel $\sigma$-algebra ${\cal B} (A)$.\note{We impose the assumption that $\Th$ and $A$ are Borel spaces to employ some of the results in Hern\'andez-Lerma and Lasserre (1996). See footnotes 9 and 11.} In addition, let $z_i^t \in \Re$ be a monetary transfer {\it from\/} player $i$ in period $t$. Given sequences $(\th^0, \th^1, \ldots)$ of type profiles and $(a^0, a^1, \ldots)$ of actions, together with $(z_i^0, z_i^1, \ldots)$ of $i$'s monetary transfers, player $i$'s total payoff is
$$\sum_{t=0}^{\infty} \d^t \Bigl(v_i(\th_i^t,a^t)-z_i^t\Bigr),$$
where (i) $\d$ is a common discount factor and $\d < 1$, and (ii) $v_i(\cdot)$ is a measurable (one-period) valuation function. The valuation function is usually called as the reward function in the Markov decision process literature. Note that we deal with the private-values environment in that player $i$'s valuation function depends only on player $i$'s type but not other players' types. Note also that we deal with the time-separable environment in that this function depends only on player $i$'s type in the current period but not other periods. We assume that $v_i(\cdot)$ is bounded, that is, $|v_i(\th_i, a)| \leq C < \infty$ for all $\th_i$ and $a$.

The dynamic evolution of players' types is represented by a stochastic kernel. Let $p(B | \th^t, a^t)$ for $B \in {\cal B} (\Th)$ be the conditional probability that the type profile lies in $B$ in period $t+1$ when the type profile is $\th^t$ and the action is $a^t$ in period $t$. We have (i) $p(\cdot| \th^t, a^t)$ is a probability measure on $\Th$ for each fixed $(\th^t, a^t)$, and (ii) $p(B | \cdot, \cdot)$ is a measurable function with respect to the product $\sigma$-algebra ${\cal B} (\Th \times A)$ for each fixed $B \in {\cal B}(\Th)$. We assume that $p(\cdot | \cdot, \cdot)$ is independent across players in the sense that $p(\th'|\th,a)= \prod_{i=1}^n p_i(\th'_i|\th_i,a)$. Observe that, except for the fact that $\th$ is private information, this environment fits into a Markov decision process with $\Th$ being the set of states.

\smallskip
\noindent {\sl 2.2. The general mechanism and the outcome efficient policy}
\medskip

We focus attention on dynamic direct mechanisms that ask each player to report his type (i.e., state) in each period and these reports are publicly observable. Let $r_i^t$ denote player $i$'s report in period $t$, which may or may not be equal to his true type $\th_i^t$. Let \par
\cl{$h_i^t =(\th_i^0, r^0, a^0, \th_i^1, r^1, a^1, \ldots, \th_i^{t-1}, r^{t-1}, a^{t-1}, \th_i^t)$}
\noindent be a private history of player $i$ in period $t$, where $r^s = (r_1^s, \ldots, r_n^s)$ for $s=0, \ldots, t-1$ is a report profile, and let $H_i^t$ be the set of all such histories. A (pure) strategy for player $i$ in period $t$ is a measurable function $\hat r_i^t: H_i^t \rightarrow \Th_i$. A strategy is truth-telling if $\hat r_i^t(h_i^t) = \th_i^t$ for all $h_i^t$. In addition, let \par
\cl{$h^t = (r^0, a^0, r^1, a^1, \ldots, r^{t-1}, a^{t-1}, r^t)$}
\noindent be a public history in period $t$ and let $H^t$ be the set of all such histories. Observe that, when players adopt the truth-telling strategy, the private histories do not contain more information than the public histories on the equilibrium path. Since we are mainly concerned with incentive compatible mechanisms in which the truth-telling strategy is an equilibrium, we will henceforth not distinguish between true states and reported states (mainly to save notations) except when explicitly stating otherwise.

In each period, the mechanism decides the action based on the actions chosen up to the previous period and the reports up to the beginning of the current period. Thus, when players adopt the truth-telling strategy, a deterministic (history-dependent) decision rule of the mechanism in period $t$ is a measurable function $\hat a^t: H^t \rightarrow A$. A special class of decision rule is the deterministic Markovian decision rule that chooses an action based only on the current state, i.e., $\hat a^t: \Th \rightarrow A$.\note{For the definition of Markovian decision rule, please refer to page 21 of Puterman (2005) or page 20 of Hern\'andez-Lerma and Lasserre (1996). It is called Markovian since it induces a Markov process over the states.}  Moreover, a randomized decision rule $\hat a^t$ specifies a probability distribution on the set of actions. Randomized decision rules may be history-dependent or Markovian. A {\it policy\/} of the mechanism is a sequence of decision rules, that is, a policy is $\pi = (\hat a^0, \hat a^1, \ldots)$. Let $\Pi$ be the set of all policies.

An outcome efficient policy is $\pi^* \in \Pi$ that maximizes the expected discounted sum of players' valuations. That is,
$$\pi^* \in \arg\max_{\pi \in \Pi} \ \ E_\th^\pi \Bigl[ \sum_{t=0}^\infty \d^{t} \sum_{j=1}^n v_j(\tilde \th_j^t, \tilde a^t) \Bigr]$$
for every $\th \in \Th$.\note{We will assume throughout that the relevant maximum is attained without specifying sufficient conditions. This assumption is valid when $A$ is compact, $v_i(\th_i, a)$ is bounded and upper-semicontinuous on  $A$ for all $\th_i$, and $p(\th' | \th, a)$ is strongly continuous, i.e., $g(\th,a) \equiv \int_{\Th} f(\th') p(d\th' | \th, a)$ is continuous and bounded on $\Th \times A$ for every measurable bounded function $f: \Th \rightarrow \Re$. Other sufficient conditions may also guarantee the existence of an outcome efficient policy $\pi^*$. See Theorem 4.2.3 of Hern\'andez-Lerma and Lasserre (1996) and the discussion preceding it.}

In addition, the mechanism specifies the monetary transfers based on public histories. A deterministic (history-dependent) transfer rule of the mechanism in period $t$ is a collection of measurable functions $\{\hat z_i^t: H^t \rightarrow \Re\}_{i \in I}$. Let $\hat z^t = (\hat z_1^t, \ldots, \hat z_n^t)$. Markovian transfer rules and randomized transfer rules can be defined similarly. In summary, a dynamic direct mechanism is represented by a family of decision rules and monetary transfer rules, $\{\hat a^t, \hat z^t\}_{t=0}^\infty$.

We call a policy {\it stationary\/} if $\hat a^t = \hat a$ for all $t$. A stationary policy has the form $\pi=(\hat a,\hat a,\ldots)$, which is denoted by $\hat a^\infty$. For the stationary environment considered in this paper,\note{The environment is stationary since both the valuation function $v_i(\cdot)$ for all $i$ and the stochastic kernel $p(\cdot | \cdot)$ do not vary with $t$.} we can restrict our attention to deterministic stationary policies when finding a policy that maximizes the expected discounted sum of players' valuations.\note{See Theorem 4.2.3 of Hern\'andez-Lerma and Lasserre (1996). Note that a deterministic stationary policy is a deterministic Markovian policy.} An outcome efficient policy thus has the form $\pi^* = (a^*)^\infty$ where $a^*: \Th \rightarrow A$. We can also restrict our attention to stationary transfer rules.

\smallskip
\noindent {\sl 2.3. The dynamic Groves mechanism}
\medskip

Define the total social welfare function $W:\Th \rightarrow \Re$ recursively by the following optimality equation (or Bellman equation):
$$W(\th)=\sum_{j=1}^n v_j(\th_j,a^*(\th))+\d \int_{\Th} W(\th') p(d\th'|\th,a^*(\th)).$$
Note that we define $W(\th)$ along an outcome efficient policy $\pi^* = (a^*)^\infty$. We can also define player $i$'s total valuation function $V_i(\th)$  recursively as
$$V_i(\th) = v_i(\th_i, a^*(\th)) + \d \int_{\Th}  V_i(\th') p(d\th'|\th,a^*(\th))$$
given $\pi^* = (a^*)^\infty$. Observe that
$$\eqalign{V_i(\th)=&v_i(\th_i, a^*(\th)) + \d \int_{\Th} v_i(\th'_i, a^*(\th')) p(d\th'|\th,a^*(\th))  \cr
+& \d^2 \int_{\Th} \int_{\Th} v_i(\th''_i, a^*(\th''))  p(d\th''|\th',a^*(\th')) p(d\th'|\th,a^*(\th)) + \cdots .}$$
Likewise, we can define the total valuation function of players other than $i$ recursively as
$$V_{-i}(\th) = \sum_{j \ne i} v_j(\th_j, a^*(\th)) + \d \int_{\Th} V_{-i}(\th') p(d\th'|\th,a^*(\th))$$
given $\pi^* = (a^*)^\infty$. Note that we use the usual notational convention that the subscript $-i$ pertains to players other than $i$. Thus, $\th_{-i} = (\th_1, \ldots, \th_{i-1}, \th_{i+1}, \ldots,\th_n)$, $\Th_{-i} = \prod_{j \ne i} \Th_j$, and so on. We now define dynamic Groves mechanisms. We note that Cavallo (2008) defined dynamic Groves mechanisms earlier.

\defn1 A dynamic Groves mechanism is a dynamic direct mechanism with an outcome efficient policy  $\pi^* = (a^*)^\infty$ and a stationary total transfer rule for player $i=1, \ldots, n$ given as
$$Z^*_i(\th) = - V_{-i}(\th) + \Phi_i(\th_{-i}).$$ \ok

Note that $\Phi_i(\cdot)$ do not depend on $\th_i$. If we recall the terminology of d'Aspremont and G\'erard-Varet (1979), the dynamic Groves mechanism is a {\it distribution mechanism\/} since the total transfer rule is given as the difference between $V_{-i}(\th)$ and the total distribution rule $\Phi_i(\th_{-i})$. In addition, the total distribution rule $\Phi_i(\th_{-i})$ is {\it discretionary\/} because it does not depend on $\th_i$.

Observe that player $i$'s total payoff in a dynamic Groves mechanism is $V_i(\th)-Z^*_i(\th) = W(\th) - \Phi_i(\th_{-i})$. Let \par
\cl{$Y_i(\th) = W(\th) - \Phi_i(\th_{-i})$.}
\noindent We can define player $i$'s (one-period) payoff $y_i(\th)$ by the identity
$$Y_i(\th) = y_i(\th) + \d \int_{\Th}  Y_i(\th') p(d\th' | \th, a^*(\th)).$$
We can also define $\Phi_i(\th_{-i})$ in terms of player $i$'s (one-period)  distribution rule $\phi_i: \Th_{-i} \rightarrow \Re$ and a given deterministic Markovian decision rule $\hat a_i: \Th_{-i} \rightarrow A$ as
$$\Phi_i(\th_{-i}) = \phi_i(\th_{-i}) + \d \int_{\Th_{-i}} \Phi_i(\th'_{-i}) p_{-i}(d\th'_{-i} | \th_{-i}, \hat a_i(\th_{-i})),$$
that is,
$$\eqalign{\Phi_i(\th_{-i})=&\phi_i(\th_{-i}) + \d \int_{\Th_{-i}} \phi_i(\th'_{-i}) p_{-i}(d\th'_{-i} | \th_{-i}, \hat a_i(\th_{-i})) \cr
+ &\d^2 \int_{\Th_{-i}} \int_{\Th_{-i}} \phi_i(\th''_{-i}) p(d\th''_{-i}|\th'_{-i}, \hat a_i(\th'_{-i})) p(d\th'_{-i} | \th_{-i}, \hat a_i(\th_{-i})) + \cdots.}$$
Thus,
$$\eqalign{&y_i(\th) = W(\th)-\Phi_i(\th_{-i}) - \d \int_{\Th} \Bigl(W(\th')-\Phi_i(\th'_{-i})\Bigr) p(d\th' | \th, a^*(\th)) \cr
= &\sum_{j=1}^n v_j(\th_j,a^*(\th))+ \d \int_{\Th} W(\th') p(d\th'|\th,a^*(\th)) \cr
- & \phi_i(\th_{-i}) - \d \int_{\Th_{-i}} \Phi_i(\th'_{-i}) p_{-i}(d\th'_{-i} | \th_{-i}, \hat a_i(\th_{-i})) \cr
- & \d \int_{\Th} W(\th') p(d\th' | \th, a^*(\th)) + \d \int_{\Th_{-i}} \Phi_i(\th'_{-i}) p_{-i}(d\th'_{-i} | \th_{-i}, a^*(\th)) \cr
= &\sum_{j=1}^n v_j(\th_j,a^*(\th))-\phi_i(\th_{-i})\cr
+ & \d\Bigl(\int_{\Th_{-i}} \Phi_i(\th'_{-i}) p_{-i}(d\th'_{-i} | \th_{-i}, a^*(\th))-\int_{\Th_{-i}} \Phi_i(\th'_{-i}) p_{-i}(d\th'_{-i} | \th_{-i}, \hat a_i(\th_{-i})) \Bigr).}$$
Then, the (one-period) monetary transfer rule $z_i^*: \Th \rightarrow \Re$ of a dynamic Groves mechanism can be defined as
$$\eqalignno{z_i^*(\th) = &v_i(\th_i,a^*(\th))-y_i(\th) \cr
= &\phi_i(\th_{-i})-\sum_{j \ne i} v_j(\th_j, a^*(\th)) \cr
+ & \d\Bigl(\int_{\Th_{-i}} \Phi_i(\th'_{-i}) p_{-i}(d\th'_{-i} | \th_{-i}, \hat a_i(\th_{-i}))-\int_{\Th_{-i}} \Phi_i(\th'_{-i}) p_{-i}(d\th'_{-i} | \th_{-i}, a^*(\th)) \Bigr).}$$
Note that the transfer $z_i^*(\th)$ depends on the report of player $i$ only through the determination of the action $a^*(\th)$, which is a prominent feature of the static Groves mechanisms. Observe that we may alternatively define dynamic Groves mechanisms using the (one-period) monetary transfer rule as follows:

\defn{1$'$} A dynamic Groves mechanism is a dynamic direct mechanism with an outcome efficient policy  $\pi^* = (a^*)^\infty$ and a stationary monetary transfer rule for player $i=1, \ldots, n$ given as
$$\eqalignno{z_i^*(\th) = &\phi_i(\th_{-i})-\sum_{j \ne i} v_j(\th_j, a^*(\th)) \cr
+ & \d\Bigl(\int_{\Th_{-i}} \Phi_i(\th'_{-i}) p_{-i}(d\th'_{-i} | \th_{-i}, \hat a_i(\th_{-i}))-\int_{\Th_{-i}} \Phi_i(\th'_{-i}) p_{-i}(d\th'_{-i} | \th_{-i}, a^*(\th)) \Bigr),}$$
where $\phi_i: \Th_{-i} \rightarrow \Re$ is player $i$'s (one-period)  distribution rule, $\hat a_i: \Th_{-i} \rightarrow A$ is a given deterministic Markovian decision rule, and $\Phi_i: \Th_{-i} \rightarrow \Re$ is defined recursively as
$$\Phi_i(\th_{-i}) = \phi_i(\th_{-i}) + \d \int_{\Th_{-i}} \Phi_i(\th'_{-i}) p_{-i}(d\th'_{-i} | \th_{-i}, \hat a_i(\th_{-i})).$$ \ok

Two special instances of dynamic Groves mechanisms are outstanding. Firstly, if $\phi_i(\th_{-i})=0$ and so $\Phi_i(\th_{-i})=0$ for all $i$ and $\th_{-i}$, then $Y_i(\th)$ becomes $W(\th)$. This mechanism is called the {\it team} mechanism by Athey and Segal (2013). Secondly, if $\hat a_i: \Th_{-i} \rightarrow A$ is given as the deterministic decision rule $a_{-i}^*: \Th_{-i} \rightarrow A$ that maximizes the expected discounted sum $E_\th^\pi [ \sum_{t=0}^\infty \d^{t} \sum_{j \ne i} v_j(\tilde \th_j^t, \tilde a^t)] $ of the valuations of players other than $i$, and $\phi_i(\th_{-i})$ is given as $\sum_{j \ne i} v_j(\th_j, a_{-i}^*(\th_{-i}))$, thus $\Phi_i(\th_{-i})$ is equal to
$$W_{-i}(\th_{-i}) = \sum_{j \ne i} v_j(\th_j, a_{-i}^*(\th_{-i})) + \d \int_{\Th_{-i}} W_{-i}(\th'_{-i}) p_{-i}(d\th'_{-i}|\th_{-i},a_{-i}^*(\th_{-i})),$$
then $Y_i(\th)$ becomes player $i$'s total marginal contribution $W(\th) - W_{-i}(\th_{-i})$.\note{Note that $W_{-i}(\th_{-i})$ is different from $V_{-i}(\th)$ defined above.} This mechanism is called the {\it dynamic pivot\/} mechanism by Bergemann and V\"alim\"aki (2010).

It is easy to establish that dynamic Groves mechanisms are periodic ex-post incentive compatible, that is, the truth-telling strategy is a best response for every player $i$ and every true type profile $\th$ in every period $t$ and private history $h_i^t$.\note{Since it is rather cumbersome to spell out the exact definition of ex-post incentive compatibility, we present it in the appendix.}

\thm1 A dynamic Groves mechanism is periodic ex-post incentive compatible. \ok

\pf By the unimprovability principle, it is sufficient to show that player $i$ does not have an incentive to `deviate now and then follow the truth-telling strategy afterwards.' Let $Y_i(r_i, \th_{-i} | \th_i)$ be player $i$'s total payoff when the true type profile is $(\th_i, \th_{-i})$ but $i$ reports $r_i$ this period. Then,\note{Note well that the transition probability $p$ depends on the (true) type profile $\th$ and the action $a$, but not directly on the report profile $r$. It depends on $r$ indirectly through $a$.}
$$\eqalign{Y(r_i, \th_{-i} | \th_i) = & \sum_{j=1}^n v_j(\th_j, a^*(r_i, \th_{-i})) - \phi_i(\th_{-i}) \cr
+ & \d \int_{\Th} W(\th') p(d\th' | \th, a^*(r_i, \th_{-i})) - \d \int_{\Th_{-i}} \Phi_i(\th'_{-i}) p_{-i}(d\th'_{-i} | \th_{-i}, \hat a_i(\th_{-i})) \cr
=& \sum_{j=1}^n v_j(\th_j, a^*(r_i, \th_{-i})) + \d \int_{\Th} W(\th') p(d\th' | \th, a^*(r_i, \th_{-i})) - \Phi_i(\th_{-i})}.$$
 Observe that $Y_i(\th_i, \th_{-i} | \th_i) = Y_i(\th) = W(\th) - \Phi_i(\th_{-i}) \geq Y_i(r_i, \th_{-i}| \th_i)$ by the definition of $W(\th)$, so a dynamic Groves mechanism is periodic ex-post incentive compatible. \endpf
\ve

\Section{The Uniqueness Results}

To establish the uniqueness of dynamic Groves mechanisms, we consider a particular class of deviations called {\it consistent deviations.\/}\note{This class of deviations is considered in Pavan {\it et al.\/} (2014), Bergemann and Strack (2015), and Es\"o and Szentes (2017). It is instrumental in rendering the dynamic mechanism design problem tractable.} Note that a deviation in a dynamic mechanism is any (reporting) strategy $\hat r_i^t: H_i^t \rightarrow \Th_i$ different from the prescribed strategy. Thus, a deviation in an incentive compatible mechanism is any strategy in which the player misreports his true type in a single or multiple periods. In a consistent deviation, after player $i$ misreports $\bar \th_i$ when his true type is $\th_i$ in this period, he keeps misreporting in all future periods. Hence, a consistent deviation is not a local deviation at one point in time, but rather represents a global deviation in the sense that the player changes his reports at every point in time. Note that the mechanism as well as other players cannot distinguish a consistent deviation from the true type realizations starting from $\bar \th_i$. That is, the same sequences of public decisions and monetary transfers are obtained. Hence, the same expectation operator is applied to the total payoffs of other players.

It is a standard fact that, given an outcome efficient policy $\pi^* = (a^*)^\infty$ where $a^*: \Th \rightarrow A$, we can describe the Markov process $\{\{\th_i^t\}_{t=0}^\infty\}_{i=1}^n$ represented by the stochastic kernel $p(\th^{t+1}|\th^t, a^*(\th^t))=\prod_{i=1}^n p_i(\th_i^{t+1} | \th_i^t, a^*(\th^t))$ alternatively as a dynamical system
$$\th_i^{t+1} = k_i(\th_i^t, a^*(\th^t), \o_i^{t+1})$$
for all $i \in N$ and $t \in \{0,1,\cdots \}$, where $k_i: \Th_i \times A \times \O_i \rightarrow \Th_i$ is a measurable mapping and $\{\o_i^t\}_{t=1}^\infty$ is a sequence of independently and identically distributed $\O_i$-valued random variables for some measurable space $\O_i$, and independent of the initial type $\th_i^0$.

\defn2 A consistent deviation is a deviation in which, after player $i$ misreports $\bar \th_i^0$ in period 0 when his true type is $\th_i^0$, he keeps misreporting in all periods as
$$\eqalign{\overline \th_i^1 &= k_i(\bar \th_i^0, a^*(\bar \th_i^0, \th_{-i}^0), \o_i^1), \cr
\bar \th_i^2 &= k_i(\bar \th_i^1, a^*(\bar \th_i^1, \th_{-i}^1), \o_i^2) \cr
&=k_i(k_i(\bar \th_i^0, a^*(\bar \th_i^0, \th_{-i}^0), \o_i^1),  a^*(k_i(\bar \th_i^0, a^*(\bar \th_i^0, \th_{-i}^0), \o_i^1), \th_{-i}^1), \o_i^2),}$$
and so on. That is, the report in period $t \in \{1, 2, \cdots\}$ is recursively given as
$$\bar \th_i^t = k_i(\bar \th_i^{t-1}, a^*(\bar \th_i^{t-1}, \th_{-i}^{t-1}), \o_i^t).$$ \ok

We are ready to define several total functions using consistent deviations. Given $\th^0 =(\th_1^0, \cdots, \th_n^0)$ and $\o = (\omega_1, \cdots, \o_n)$ where $\o_i =\{\o_i^t\}_{t=1}^\infty$, let $\th_i^1=k_i(\th_i^0, a^*(\th_i^0, \th_{-i}^0), \o_i^1), \bar \th_i^1=k_i(\bar \th_i^0, a^*(\bar \th_i^0, \th_{-i}^0), \o_i^1), \th_i^2=k_i(\th_i^1, a^*(\th_i^1, \th_{-i}^1), \o_i^2), \bar \th_i^2=k_i(\bar \th_i^1,$ $a^*(\bar \th_i^1, \th_{-i}^1), \o_i^2)$, and so on for all $\th_i^t$ and $\bar \th_i^t$ for $t \in \{1, 2, \cdots\}$. Define
$$V_i^D(\th_i^0, \th_{-i}^0, a^*(\bar \th_i^0, \th_{-i}^0),\o)=\sum_{t=0}^\infty \d^t v_i(\th_i^t,a^*(\bar \th_i^t, \th_{-i}^t))$$
and
$$V_i^C(\th_i^0, \th_{-i}^0, a^*(\bar \th_i^0, \th_{-i}^0))=E\bigl[\sum_{t=0}^\infty \d^t v_i(\th_i^t,a^*(\bar \th_i^t, \th_{-i}^t))\bigr]$$
where the expectation is taken over $\o$. We also define
$$V_{-i}^D(\th_{-i}^0, a^*(\bar \th_i^0, \th_{-i}^0),\o)=\sum_{t=0}^\infty \d^t \sum_{j \ne i} v_j(\th_j^t,a^*(\bar \th_i^t, \th_{-i}^t)),$$
$$V_{-i}^C(\th_{-i}^0, a^*(\bar \th_i^0, \th_{-i}^0))=E\bigl[\sum_{t=0}^\infty \d^t \sum_{j \ne i} v_j(\th_j^t,a^*(\bar \th_i^t, \th_{-i}^t))\bigr],$$
$$Z_i^D(\bar \th_i^0, \th_{-i}^0, \o)=\sum_{t=0}^\infty \d^t z_i(\bar \th_i^t, \th_{-i}^t) {\rm \ \ and \ \ } Z_i^C(\bar \th_i^0, \th_{-i}^0) = E\bigl[ Z_i^D(\bar \th_i^0, \th_{-i}^0) \bigr].$$
Note that these functions do not depend on $\th_i^0$. Define
$$U_i^C(\th_i^0, \th_{-i}^0, a^*(\bar \th_i^0, \th_{-i}^0))=V_i^C(\th_i^0, \th_{-i}^0, a^*(\bar \th_i^0, \th_{-i}^0))-Z_i^C(\bar \th_i^0, \th_{-i}^0)$$
and
$$W^C(\th_i^0, \th_{-i}^0,a^*(\bar \th_i^0, \th_{-i}^0))=V_i^C(\th_i^0, \th_{-i}^0, a^*(\bar \th_i^0, \th_{-i}^0))+V_{-i}^C(\th_{-i}^0, a^*(\bar \th_i^0, \th_{-i}^0)).$$
Observe that, since the environment is stationary and Markov, it does not matter whether the period begins in $t=0$ or any $t=0, 1, \cdots$. Hence, we will drop the superscript for $t=0$ and write $V_i^C(\th_i, \th_{-i}, a^*(\bar \th_i, \th_{-i}))$ and so on. Observe also that $V_i^C(\th_i, \th_{-i}, a^*(\th_i, \th_{-i}))=V_i(\th_i, \th_{-i})$ where the latter is defined in Section 2. Likewise, $V_{-i}^C(\th_{-i}, a^*(\th_i, \th_{-i}))=V_{-i}(\th_i, \th_{-i})$ and $W^C(\th_i, \th_{-i}, a^*(\th_i, \th_{-i}))=W(\th_i, \th_{-i})$. Let $Z_i(\th_i, \th_{-i})=Z_i^C(\th_i, \th_{-i})$ and $U_i(\th_i, \th_{-i})=V_i(\th_i, \th_{-i})-Z_i(\th_i, \th_{-i})$.

Having defined these total functions, we henceforth follow Carbajal (2010) as closely as possible to demonstrate that many of the results for the dynamic setting can be obtained by porting the corresponding results of the static mechanism design. We need additional assumptions. First, assume that $\Th_i$ is an open connected subset of $\Re^{k_{i}}$. Next, assume that the domain ${\cal V}_i$ of player $i$'s total valuations consists of $V_i^C(\th_i, \th_{-i}, a^*(\bar \th_i, \th_{-i}))$'s that are equi-Lipschitz continuous and regular on $\Th_i$.

A family of functions $\{V_i^C(\th_i, \th_{-i}, a^*(\bar \th_i, \th_{-i})): \Th_i \rightarrow \Re | \th_{-i} \in \Th_{-i}, \bar \th_i \in \Th_i \}$ is equi-Lipschitz continuous on $\Th_i$ if there exists a non-negative number $L_i$ such that \par
\cl{$| V_i^C(\th_i, \th_{-i}, a^*(\bar \th_i, \th_{-i})) - V_i^C(\hat \th_i, \th_{-i}, a^*(\bar \th_i, \th_{-i}))| \leq L_i ||\th_i - \hat \th_i||$}
\noindent for all $\th_i, \hat \th_i$, $\th_{-i}$, and $\bar \th_i$. As for regularity, given an open set $Y \subseteq \Re^k$ and a function $g$ on $Y$ to $\Re$, the one-sided directional derivative of $g$ at $y \in Y$ in the direction of $d \in \Re^k$ is defined as
$D^{+} g(y;d) = \lim_{\lambda \downarrow 0} \  [g(y+\lambda d) - g(y)]/\lambda,$
provided this limit exists. The function $g$ is regular at $y \in Y$ if it admits one-sided directional derivatives at $y$ in any direction $d$, and $g$ is regular on $Y$ if it is regular at every $y \in Y$. Please refer to Carbajal (2010) for a more detailed discussion of these concepts.
These assumptions in particular imply that the following limits exist and are finite:\note{Observe that $D^{-}_{\th_{i}} V_i^C(\th_i,\th_{-i}, a^*(\bar \th_i, \th_{-i});d) = -D^{+}_{\th_{i}} V_i^C(\th_i,\th_{-i}, a^*(\bar \th_i, \th_{-i});-d)$.}

{\smalltype
$$\eqalign{&D^{+}_{\th_{i}} V_i^C(\th_i, \th_{-i},a^*(\bar \th_i, \th_{-i});d) = \lim_{\lambda \downarrow 0} \frac{V_i^C(\th_i+\lambda d, \th_{-i}, a^*(\bar \th_i, \th_{-i})) - V_i^C(\th_i, \th_{-i}, a^*(\bar \th_i, \th_{-i}))}{\lambda}, \ \forall d \in \Re^{k_{i}}, \cr
&D^{-}_{\th_{i}} V_i^C(\th_i, \th_{-i},a^*(\bar \th_i, \th_{-i});d) = \lim_{\lambda \uparrow 0} \frac{V_i^C(\th_i+\lambda d, \th_{-i}, a^*(\bar \th_i, \th_{-i})) - V_i^C(\th_i, \th_{-i}, a^*(\bar \th_i, \th_{-i}))}{\lambda}, \ \forall d \in \Re^{k_{i}}.}$$
}

We note that the conditions on $V_i^C(\th_i, \th_{-i}, a^*(\bar \th_i, \th_{-i}))$ are imposed only with respect to the outcome efficient decision rule $a^*(\cdot)$, not with respect to any possible decision rule $\hat a(\cdot)$. We also note that the conditions on $V_i^C(\th_i, \th_{-i}, a^*(\bar \th_i, \th_{-i}))$ can be passed over to the conditions on $v_i(\th_i, a)$ and $p(B|\th_i, \th_{-i}, a)$. For Lipschitz continuity of $V_i^C(\th_i, \th_{-i},  a^*(\bar \th_i,$ $\th_{-i}))$, it is sufficient to assume that both $v_i(\th_i, a)$ and  $p(B|\th_i, \th_{-i}, a)$ are Lipschitz continuous on $\Th_i$.\note{See, for instance, Dufour and Prieto-Rumeau (2012) for Lipschitz continuity of a stochastic kernel.} Similarly, for regularity of $V_i^C(\th_i, \th_{-i}, a^*(\bar \th_i, \th_{-i}))$, it is sufficient to assume that $v_i(\th_i, a)$ is regular on $\Th_i$ and that, for any $\th \in \Th$ and any direction $d \in \prod_{i=1}^n \Re^{k_{i}}$,
$$\lim_{\lambda \downarrow 0} \frac{\int_{\Th} w(\th', a) p(d\th' | \th + \lambda d, a) - \int_{\Th} w(\th',a) p(d\th' | \th, a)  }{\lambda}$$
exists and is finite for any function $w(\th, a)$. Other conditions on $v_i(\th_i, a)$ and $p(B|\th_i, \th_{-i}, a)$ may also lead us to the desired conditions on $V_i^C(\th_i, \th_{-i},a^*(\bar \th_i, \th_{-i}))$. We first prove two straightforward lemmas.

\lemma1 Assume that $\Th_i$ is an open connected subset of $\Re^{k_{i}}$ and that the domain ${\cal V}_i$ of player $i$'s total valuations consists of $V_i^C(\th_i,\th_{-i}, a^*(\bar \th_i, \th_{-i}))$'s that are equi-Lipschitz continuous and regular on $\Th_i$. If a dynamic direct mechanism with an outcome efficient policy $\pi^* = (a^*)^\infty$ and a stationary total transfer rule $Z_i: \Th \rightarrow \Re$ is periodic ex-post incentive compatible, then $U_i(\th_i, \th_{-i})$ and $W(\th_i, \th_{-i})$ are Lipschitz continuous and differentiable almost everywhere on $\Th_i$. \ok

\pf See the appendix. \endpf

\lemma2 Assume that $\Th_i$ is an open connected subset of $\Re^{k_{i}}$ and that the domain ${\cal V}_i$ of player $i$'s total valuations consists of $V_i^C(\th_i,\th_{-i}, a^*(\bar \th_i, \th_{-i}))$'s that are equi-Lipschitz continuous and regular on $\Th_i$. Let $(\pi^*,\{Z_i\}_{i=1}^n)$ be an outcome efficient and periodic ex-post incentive compatible dynamic direct mechanism. Given any $\th_{-i} \in \Th_{-i}$, if $W(\th_i,\th_{-i})$ is regular at $\th_i \in \Th_i$, then for any direction $d \in \Re^{k_{i}}$ we have
$$\eqalign{&D^{+}_{\th_{i}} V_i^C(\th_i,\th_{-i}, a^*(\th_i, \th_{-i});d) \leq D^{+}_{\th_{i}} W(\th_i, \th_{-i};d), \cr
&D^{-}_{\th_{i}} V_i^C(\th_i,\th_{-i}, a^*(\th_i, \th_{-i});d) \geq D^{-}_{\th_{i}} W(\th_i, \th_{-i};d).}$$ \ok

\pf See the appendix. \endpf

An immediate consequence of this lemma is that, if $W(\th_i, \th_{-i})$ is differentiable at $\th_i$ so that $D^{+}_{\th_{i}} W(\th_i, \th_{-i};d)= D^{-}_{\th_{i}} W(\th_i,\th_{-i};d)$ at $\th_i$, then
$$D^{-}_{\th_{i}} V_i^C(\th_i,\th_{-i}, a^*(\th_i, \th_{-i});d) \geq D^{+}_{\th_{i}} V_i^C(\th_i,\th_{-i}, a^*(\th_i, \th_{-i});d).$$

\noindent The reverse of this inequality is stated as a property.

\defn3 A dynamic direct mechanism with an outcome efficient policy $\pi^* = (a^*)^\infty$ and a stationary total transfer rule $Z_i: \Th \rightarrow \Re$ satisfies Property A if, for every $i$, every $\th_{-i}$, and each $\th_i$ such that $W(\th_i, \th_{-i})$ is differentiable at $\th_i$, we have
$$D^{-}_{\th_{i}} V_i^C(\th_i,\th_{-i}, a^*(\th_i, \th_{-i});d) \leq D^{+}_{\th_{i}} V_i^C(\th_i,\th_{-i}, a^*(\th_i, \th_{-i});d)$$ \ok

\noindent for any direction $d \in \Re^{k_{i}}$.

\noindent {\sl Note:} This property corresponds to Property 1 of Carbajal (2010).

When this property is satisfied, Lemma 2 implies that $D^{-}_{\th_{i}} V_i^C(\th_i,\th_{-i}, a^*(\th_i, \th_{-i});d) = D^{+}_{\th_{i}} V_i^C(\th_i,\th_{-i}, a^*(\th_i, \th_{-i});d)$ for any direction $d \in \Re^{k_{i}}$ when $W(\th_i, \th_{-i})$ is differentiable. Hence, $V_i^C(\th_i, \th_{-i}, a^*(\th_i, \th_{-i}))$ admits two-sided derivatives with respect to $\th_i$, which is key for the uniqueness result.

\thm2 Assume that $\Th_i$ is an open connected subset of $\Re^{k_{i}}$ and that the domain ${\cal V}_i$ of player $i$'s total valuations consists of $V_i^C(\th_i,\th_{-i}, a^*(\bar \th_i, \th_{-i}))$'s that are equi-Lipschitz continuous and regular on $\Th_i$. Then, any dynamic direct mechanism which is outcome efficient and periodic ex-post incentive compatible is a dynamic Groves mechanism if and only if it satisfies Property A. \ok

\pf For sufficiency, assume that Property A is satisfied, and let $(\pi^*, \{Z_i\}_{i=1}^n)$ be an outcome efficient and periodic ex-post incentive compatible dynamic direct mechanism. Define $\Phi_i^C(\bar \th_i,\th_{-i}) = Z_i^C(\bar \th_i, \th_{-i})+V_{-i}^C(\th_{-i},a^*(\bar \th_i, \th_{-i}))$ for this mechanism. Define also that $\Phi_i(\th_i, \th_{-i})=\Phi_i^C(\th_i, \th_{-i})$. It suffices to show that $\Phi_i^C(\th_i, \th_{-i}) = \Phi_i(\th_i, \th_{-i})$ is constant over $\Th_i$.

Fix $\th_{-i} \in \Th_{-i}$. Since $U_i(\th)= V_i(\th)-Z_i(\th) = V_i(\th) + V_{-i}(\th)-\Phi_i^C(\th) = W(\th) - \Phi_i^C(\th)$, we have $\Phi_i^C(\th) = W(\th)-U_i(\th)$ and so $\Phi_i^C(\cdot, \th_{-i})$ is Lipschitz continuous and differentiable almost everywhere on $\Th_i$ by Lemma 1. We claim that for each direction $d \in \Re^{k_{i}}$, the two-sided directional derivative of $\Phi_i^C(\cdot, \th_{-i})$ in the direction of $d$, denoted by $D_{\th_{i}} \Phi_i^C(\cdot, \th_{-i};d)$, is zero a.e. on $\Th_i$, from which it follows that $\Phi_i^C(\cdot, \th_{-i})$ is constant over $\Th_i$. To see this, fix $d \in \Re^{k_{i}}$ and define, for each $\th_i \in \Th_i$, the auxiliary functions $\psi_s$ and $\psi_i$ on $\Re$ by
$$\eqalign{\psi_s(\lambda) = &V_i^C(\th_i+\lambda d, \th_{-i}, a^*(\th_i, \th_{-i})) + V_{-i}^C(\th_{-i}, a^*(\th_i, \th_{-i})) - W(\th_i+\lambda d, \th_{-i}),\cr
\psi_i(\lambda) = &V_i^C(\th_i+\lambda d, \th_{-i}, a^*(\th_i, \th_{-i})) + V_{-i}^C(\th_{-i}, a^*(\th_i, \th_{-i})) - \Phi_i^C(\th_i, \th_{-i}) \cr
- & U_i(\th_i + \lambda d, \th_{-i}).}$$
\noindent Note that $\psi_s(\lambda) \leq 0$ for any $\lambda$ by definition of the total social welfare function $W:\Th \rightarrow \Re$ and $\psi_s(0)=0$. Similarly, $\psi_i(\lambda) \leq 0$ for any $\lambda$ by periodic ex-post incentive compatibility and $\psi_i(0)=0$.

Suppose $\th_i$ is a type in $\Th_i$ at which both $W(\cdot, \th_{-i})$ and $U_i(\cdot, \th_{-i})$ are differentiable. Then, Property A implies that
$$\eqalign{0 \leq &\lim_{\lambda \uparrow 0} \frac{\psi_s(\lambda)-\psi_s(0)}{\lambda} = D^{-}_{\th_{i}} V_i^C(\th_i, \th_{-i}, a^*(\th_i, \th_{-i});d) - D^{-}_{\th_{i}} W(\th_i, \th_{-i};d) \cr
\leq &D^{+}_{\th_{i}} V_i^C(\th_i, \th_{-i}, a^*(\th_i, \th_{-i});d)- D^{+}_{\th_{i}} W(\th_i, \th_{-i};d) = \lim_{\lambda \downarrow 0} \frac{\psi_s(\lambda)-\psi_s(0)}{\lambda} \leq 0.}$$
\noindent A similar argument holds if we use $\psi_i$ above instead. It follows that, for almost every type $\th_i \in \Th_i$, we have $\lim_{\lambda \rightarrow 0}\psi_s(\lambda)/\lambda = \lim_{\lambda \rightarrow 0}\psi_i(\lambda)/\lambda = 0$. Hence, a.e. on $\Th_i$,
$$D_{\th_{i}} \Phi_i^C(\th_i, \th_{-i};d) = \lim_{\lambda \rightarrow 0} \frac{\Phi_i^C(\th_i+\lambda d, \th_{-i})-\Phi_i^C(\th_i, \th_{-i})}{\lambda} = \lim_{\lambda \rightarrow 0} \frac{\psi_i(\lambda)-\psi_s(\lambda)}{\lambda} = 0.$$
Thus, for any direction $d \in \Re^{k_{i}}$, we have $D_{\th_{i}} \Phi_i^C(\th_i, \th_{-i};d)=0$ almost everywhere on $\Th_i$ as claimed.

For necessity, assume that any dynamic direct mechanism which is outcome efficient and periodic ex-post incentive compatible is a dynamic Groves mechanism. Fix an outcome efficient and periodic ex-post incentive compatible dynamic direct mechanism $(\pi^*, \{Z_i\}_{i=1}^n)$ and $\th_{-i} \in \Th_{-i}$. Since this mechanism is a dynamic Groves mechanism, $\Phi_i(\cdot, \th_{-i})$ is constant over $\Th_i$. Hence, for any $\th_i \in \Th_i$ and any direction $d \in \Re^{k_{i}}$, its two-sided directional derivative vanishes, i.e., $D_{\th_{i}} \Phi_i(\th_i, \th_{-i}; d)=0$. With the auxiliary function $\psi_s$ and $\psi_i$ defined above, we have
$$\eqalign{0 = &\lim_{\lambda \uparrow 0} \frac{\Phi_i(\th_i + \lambda d,\th_{-i})-\Phi_i(\th_i, \th_{-i})}{\lambda} \cr
= &\lim_{\lambda \uparrow 0} \frac{W(\th_i+\lambda d,\th_{-i})-U_i(\th_i+\lambda d,\th_{-i})-\Phi_i(\th_i, \th_{-i})}{\lambda} \cr
= &\lim_{\lambda \uparrow 0} \frac{\psi_i(\lambda)-\psi_s(\lambda)}{\lambda} = \lim_{\lambda \uparrow 0} \frac{[\psi_i(\lambda)-\psi_i(0)]-[\psi_s(\lambda)-\psi_s(0)]}{\lambda}.}$$
A similar argument holds if we let $\lambda$ approach zero from above. These facts, together with the regularity of the auxiliary functions at any $\th_i$ where $W(\cdot, \th_{-i})$ and $U_i(\cdot, \th_{-i})$ admit derivatives, imply that
$$D^{-}\psi_i(0)-D^{-}\psi_s(0) = D^{+}\psi_i(0)-D^{+}\psi_s(0).$$

Observe now that $\psi_i(0)=W(\th_i, \th_{-i})-U_i(\th_i,\th_{-i})-\Phi_i(\th_i,\th_{-i})$. Thus, if $W(\cdot, \th_{-i})$ and $U_i(\cdot, \th_{-i})$ are differentiable at $\th_i$, it follows that $\psi_i$ is differentiable at 0, and hence $D^{-}\psi_i(0)=D^{+}\psi_i(0)$. Then, from the equation above, $\psi_s$ is also differentiable at 0. Using the definition of $\psi_s$, we know that the function $V_i^C(\cdot, \th_{-i}, a^*(\th_i, \th_{-i}))$ admits the two-sided directional derivative at $\th_i$ in the direction of $d \in \Re^{k_{i}}$. Thus, Property A is satisfied. \endpf

This theorem corresponds to Theorem 1 of Carbajal (2010). Additionally, we can provide a sufficient condition for the uniqueness of dynamic Groves mechanisms on restricted domains, which corresponds to Corollary 1 of Carbajal (2010). Note that the family $\{V_i^C(\cdot, \th_{-i}, a^*(\bar \th_i, \th_{-i})): \Th_i \rightarrow \Re | \th_{-i} \in \Th_{-i}, \bar \th_i \in \Th_i \}$ is said to be pointwise bounded on $\Th_i$ if, for each $\th_i \in \Th_i$, the set of real numbers $\{V_i^C(\th_i, \th_{-i}, a^*(\bar \th_i, \th_{-i}))\}_{ \th_{-i} \in \Th_{-i}, \bar \th_i \in \Th_i}$ is bounded.

\corr1 Assume that $\Th_i$ is an open, convex bounded subset of $\Re^{k_{i}}$ and that the domain of player $i$'s total valuations $\{V_i^C(\cdot, \th_{-i}, a^*(\bar \th_i, \th_{-i})): \Th_i \rightarrow \Re | \th_{-i} \in \Th_{-i}, \bar \th_i \in \Th_i \}$ is a collection of pointwise bounded, convex functions on $\Th_i$. Then, any dynamic direct mechanism which is outcome efficient and periodic ex-post incentive compatible is a dynamic Groves mechanism. \ok

\pf We omit the proof since it is almost identical to that in Carbajal (2010). \endpf

As Holmstr\"om (1979) states, it is conceivable that uniqueness would be lost when the domain is restricted. Compared to Cavallo's (2008) result for the unrestricted domain, this corollary shows the uniqueness result for a restricted domain.\note{Cavallo follows the method of proof in Green and Laffont (1977), which rests crucially on the assumption of a large domain of valuations to establish the uniqueness.} Observe that in many interesting economic applications, including the repeated auctions and nonlinear pricing models, the domain of valuations is indeed restricted and further satisfies the conditions of Corollary 1.\note{We refer the reader to page 1115 of Krishna and Maenner (2001) for this observation.} We illustrate this point with a simple nonlinear pricing example.

\eg1 There is a single player who is interested in a good that a monopolist produces. Since there is only one player, we drop the subscript in this example for notational convenience. Let $\Th=(0,1)$ and $A=[0,1]$. Note that we have a one-dimensional type space in this example. The (one-period) valuation function is given as $v(\th,a)=\th a$, where $a$ is the probability that the player gets the good. Let $c$ denote the constant marginal cost of producing the good. Thus, an efficient decision rule is such that $a^*(\th)=1$ when $\th \geq c$ and $a^*(\th)=0$ otherwise. The transition kernel is given as follows: Let $\o = \{\o^t\}_{t=1}^\infty$ be a sequence of independently and identically distributed random variables on $(-1,1)$ and let
$$\th^{t+1}=\cases{\g \th^t + (1-\g) \o^{t+1} & if $0 < \th^{t+1} < 1$; \cr
                    \g \th^t + (1-\g) \o^{t+1} - 1 & if $\th^{t+1} > 1$; \cr
                    \g \th^t + (1-\g) \o^{t+1} + 1 & if $\th^{t+1} < 0$,
                   }$$
where $\g$ is a real number in $(0,1)$.\note{We can assign $\th^{t+1}$ to any value in $(0,1)$ for the measure-zero event of $\th^{t+1}=0$ or $1$.} We have $\partial \th^t/\partial \th^0 = \g^t$ and $\partial V^D(\th^0, a^*(\bar \th^0), \o)/\partial \th^0 = \sum_{t=0}^\infty \d^t \g^t a^*(\bar \th^t)$. Hence, $V^C(\th^0, a^*(\bar \th^0))$ is linear in $\th^0$, as well as other conditions of Corollary 1 are satisfied.\note{Note that $\th^t$'s are independent across periods when $\g=0$ and perfectly correlated when $\g=1$. It is easy to see that  $V^C(\th^0, a^*(\bar \th^0))$ is linear in $\th^0$ as well for these cases.} \ok

Theorem 2 also immediately implies that the uniqueness result holds when the domain of player $i$'s total valuations $\{V_i^C(\cdot, \th_{-i}, a^*(\bar \th_i, \th_{-i})): \Th_i \rightarrow \Re | \th_{-i} \in \Th_{-i}, \bar \th_i \in \Th_i \}$ is a collection of pointwise bounded, continuously differentiable functions on an open connected subset $\Th_i$ of $\Re^{k_{i}}$.

\Section{Conclusion}

With a careful specification of dynamic environments, we have examined necessary and sufficient conditions for the uniqueness of dynamic Groves mechanisms. We first assumed that the set $\Th_i$ of types and the set $A$ of actions are Borel spaces and showed that dynamic Groves mechanisms are outcome efficient and periodic ex-post incentive compatible. Next, with the additional assumptions that $\Th_i$ is an open connected subset of $\Re^{k_{i}}$ and the domain ${\cal V}_i$ of player $i$'s total valuations consists of $V_i^C(\th_i,\th_{-i},a^*(\bar \th_i, \th_{-i}))$'s that are equi-Lipschitz continuous and regular on $\Th_i$, we have provided necessary and sufficient conditions for the uniqueness of dynamic Groves mechanisms. This is an extension of the results in Carbajal (2010) to the dynamic setting.

We have obtained the uniqueness result for stationary Markovian environments. We have utilized the recursive structure and also employed the results in the Markov decision process literature, in particular, for the existence and sufficiency of deterministic stationary policies. It is a future research agenda to extend the uniqueness result to more general (i.e., non-stationary non-Markovian) dynamic environments.
`
\appx

\noindent {\bf Definition of ex-post incentive compatibility:} A mechanism $\{\hat a^t, \hat z^t\}_{t=0}^\infty$ is periodic ex-post incentive compatible if $\forall i, \forall t, \forall h^{t-1}, \forall a^{t-1}, \forall (\th_i^t, \th_{-i}^t)$, and $\forall r_i^t$:
$$\eqalign{&v_i(\th_i^t, \hat a^t(h^{t-1},a^{t-1},\th_i^t, \th_{-i}^t))- \hat z_i^t(h^{t-1},a^{t-1}, \th_i^t, \th_{-i}^t) \cr
+ &\d \int_{\Th} \bigl(v_i(\th_i^{t+1}, \hat a^{t+1}(h^{t+1})) - \hat z_i^{t+1}(h^{t+1}) \bigr)p(d\th^{t+1}| \th^t, \hat a^t(h^{t-1},a^{t-1},\th_i^t, \th_{-i}^t)) \cr
+ &\d^2 \int_{\Th} \int_{\Th} \bigl(v_i(\th_i^{t+2}, \hat a^{t+2}(h^{t+2}))- z_i^{t+2}(h^{t+2}) \bigr) p(d\th^{t+2}|\th^{t+1},\hat a^{t+1}(h^{t+1})) \cr
& \times p(d\th^{t+1}|\th^t, \hat a^t(h^{t-1},a^{t-1},\th_i^t, \th_{-i}^t)) + \cdots }$$
$$\eqalign{\geq &v_i(\th_i^t, \hat a^t(h^{t-1},a^{t-1},r_i^t, \th_{-i}^t))- \hat z_i^t(h^{t-1},a^{t-1}, r_i^t, \th_{-i}^t) \cr
+ &\d \int_{\Th} \bigl(v_i(\th_i^{t+1}, \hat a^{t+1}(\overline h^{t+1})) - \hat z_i^{t+1}(\overline h^{t+1}) \bigr)p(d\th^{t+1}| \th^t, \hat a^t(h^{t-1},a^{t-1},r_i^t, \th_{-i}^t)) \cr
+ &\d^2 \int_{\Th} \int_{\Th} \bigl(v_i(\th_i^{t+2}, \hat a^{t+2}(\overline h^{t+2}))- z_i^{t+2}(\overline h^{t+2}) \bigr) p(d\th^{t+2}|\th^{t+1},\hat a^{t+1}(\overline h^{t+1})) \cr
& \times p(d\th^{t+1}|\th^t, \hat a^t(h^{t-1},a^{t-1}, r_i^t, \th_{-i}^t)) + \cdots, }$$
where we define
$$\eqalign{&h^{t+1}=(h^{t-1}, a^{t-1}, \th_i^t, \th_{-i}^t, \hat a^t(h^{t-1}, a^{t-1}, \th_i^t, \th_{-i}^t), \th^{t+1}); h^{t+2}=(h^{t+1},\hat a^{t+1}(h^{t+1}), \th^{t+2});\cr
&\overline h^{t+1}=(h^{t-1}, a^{t-1}, r_i^t, \th_{-i}^t, \hat a^t(h^{t-1}, a^{t-1}, r_i^t, \th_{-i}^t), \th^{t+1}); \overline h^{t+2}=(\overline h^{t+1}, \hat a^{t+1}(\overline h^{t+1}),\th^{t+2}).}$$

\pfo{Lemma 1} For any two distinct $\th_i$ and $\hat \th_i$ in $\Th_i$, we have
$$\eqalign{&U_i(\th_i, \th_{-i})-U_i(\hat \th_i, \th_{-i}) \leq U_i^C(\th_i, \th_{-i}, a^*(\th_i, \th_{-i}))-U_i^C(\hat \th_i, \th_{-i}, a^*(\th_i, \th_{-i})) \cr
= &V_i^C(\th_i, \th_{-i}, a^*(\th_i, \th_{-i}))-V_i^C(\hat \th_i, \th_{-i}, a^*(\th_i, \th_{-i})) \leq L_i ||\th_i - \hat \th_i||}$$
where the first inequality follows from periodic ex-post incentive compatibility and the second inequality follows from equi-Lipschitz continuity. Reversing the roles of $\th_i$ and $\hat \th_i$, we have $|U_i(\th_i, \th_{-i}) - U_i(\hat \th_i, \th_{-i})| \leq L_i ||\th_i - \hat \th_i||$. Hence, $U_i(\th_i, \th_{-i})$ is Lipschitz continuous on $\Th_i$.  A similar argument holds for $W(\th_i, \th_{-i})$. Being Lipschitz functions defined on an open connected set $\Th_i \subseteq \Re^{k_{i}}$, the functions $U_i$ and $W$ are differentiable almost everywhere on $\Th_i$. \endpf

\pfo{Lemma 2} Given $\th_{-i}$, we have
$$\eqalign{&W(\th_i+\lambda d, \th_{-i})-W(\th_i, \th_{-i}) \cr
\geq & V_i^C(\th_i+\lambda d, \th_{-i}, a^*(\th_i, \th_{-i}))+V_{-i}^C(\th_{-i}, a^*(\th_i, \th_{-i})) \cr
-&V_i^C(\th_i, \th_{-i}, a^*(\th_i,\th_{-i}))-V_{-i}^C(\th_{-i},a^*(\th_i,\th_{-i})) \cr
=& V_i^C(\th_i+\lambda d, \th_{-i}, a^*(\th_i, \th_{-i}))-V_i^C(\th_i, \th_{-i}, a^*(\th_i,\th_{-i}))}$$
where the inequality follows from the definition of the total social welfare function $W: \Th \rightarrow \Re$.
Thus, if $\lambda > 0$ we have
$$\frac{V_i^C(\th_i+\lambda d, \th_{-i}, a^*(\th_i, \th_{-i}))-V_i^C(\th_i, \th_{-i}, a^*(\th_i, \th_{-i}))}{\lambda}
\leq \frac{W(\th_i+\lambda d, \th_{-i})-W(\th_i, \th_{-i})}{\lambda},$$
whereas if $\lambda < 0$ we have
$$\frac{V_i^C(\th_i+\lambda d, \th_{-i}, a^*(\th_i, \th_{-i}))-V_i^C(\th_i, \th_{-i}, a^*(\th_i, \th_{-i}))}{\lambda} \geq \frac{W(\th_i+\lambda d, \th_{-i})-W(\th_i, \th_{-i})}{\lambda}.$$
If $W(\th_i, \th_{-i})$ is regular at $\th_i$, we get the desired results as we let $\lambda \downarrow 0$ and $\lambda \uparrow 0$ respectively. \endpf

\ref

\paper{Athey, S. and Segal, I.}{2013}{An efficient dynamic mechanism}{\emet 81}{2463-2485}

\paper{Bergemann, D. and Said, M.}{2010}{Dynamic auctions}{Cochran, J., Cox, L., Keskinocak, P., Kharoufeh, J., and Smith, C. (Eds.), {\it Wiley Encyclopedia of Operations Research and Management Science,\/} Wiley}{1511-1522}

\paper{Bergemann, D. and Strack, P.}{2015}{Dynamic revenue maximization: A continuous time approach}{\jet 159}{819-853}

\paper{Bergemann, D. and V\"alim\"aki, J.}{2010}{The dynamic pivot mechanism}{\emet 78}{771-789}

\paper{Bergemann, D. and V\"alim\"aki, J.}{2019}{Dynamic mechanism design: An introduction}{{\it Journal of Economic Literature\/} 57}{235-274}

\paper{Carbajal, J. C.}{2010}{On the uniqueness of Groves mechanisms and the payoff equivalence principle}{\geb 68}{763-772}

\paper{Cavallo, R.}{2008}{Efficiency and redistribution in dynamic mechanism design}{{\it Proceedings of the 9th ACM Conference on Electronic Commerce\/}}{220-229}

\paper{d'Aspremont, C. and G\'erard-Varet, L.-A.}{1979}{Incentives and incomplete information}{\jpube 11}{25-45}

\paper{Dufour, F. and Prieto-Rumeau, T.}{2012}{Approximation of infinite horizon discounted cost Markov decision processes}{Hern\'andez-Hern\'andez, D. and Minj\'arez-Sosa, J. A. (Eds.), {\it Optimization, Control, and Applications of Stochastic Systems,\/} Springer}{59-76}

\paper{Es\"o, P. and Szentes, B.}{2017}{Dynamic contracting: An irrelevance theorem}{\te 12}{109-139}

\paper{Green, J. and Laffont, J.-J.}{1977}{Characterization of satisfactory mechanisms for the revelation of preferences for public goods}{\emet 45}{427-438}

\book{Hern\'andez-Lerma, O. and Lasserre, J.}{1996}{Discrete-Time Markov Control Processes: Basic Optimality Criteria}{Springer}

\paper{Holmstr\"om, B.}{1979}{Groves' scheme on restricted domains}{\emet 47}{1137-1144}

\paper{Krishna, V. and Maenner, E.}{2001}{Convex potentials with an application to mechanism design}{\emet 69}{1113-1119}

\wp{Krishna, V. and Perry, M.}{2000}{Efficient mechanism design}{manuscript}

\paper{Milgrom, P. and Segal, I.}{2002}{Envelope theorems for arbitrary choice sets}{\emet 70}{583-601}

\paper{Pavan, A.}{2017}{Dynamic mechanism design: Robustness and endogenous types}{Honor\'e, B, Pakes, A., Piazzesi, M., and Samuelson, L. (Eds.), {\it Advances in Economics and Econometrics: Eleventh World Congress,\/} Cambridge University Press}{1-62}

\paper{Pavan, A., Segal, I. and Toikka, J.}{2014}{Dynamic mechanism design: A Myersonian approach}{\emet 82}{601-653}

\book{Puterman, M.}{2005}{Markov Decision Processes: Discrete Stochastic Dynamic Programming}{John Wiley \& Sons}

\paper{Skrzypacz, A. and Toikka, J.}{2015}{Mechanisms for repeated trade}{{\it American Economic Journal: Microeconomics\/} 7}{252-293}

\paper{Vohra, R.}{2012}{Dynamic mechanism design}{{\it Surveys in Operations Research and Management Science} 17}{60-68}

\paper{Walker, M}{1978}{A note on the characterization of mechanisms for the revelation of preferences}{\emet 46}{147-152}

\paper{Williams, S.}{1999}{A characterization of efficient, bayesian incentive compatible mechanisms}{\et 14}{155-180}

\bye